\documentclass[a4paper]{jpconf}
\usepackage{graphicx}
\begin{document}
\title{Magnetoelectric Properties of (Ca$_{1-x}$Sr$_{x}$)$_2$CoSi$_2$O$_7$ Crystals}

\author{M. Akaki, J. Tozawa, M. Hitomi, D. Akahoshi, and H. Kuwahara}

\address{Department of Physics, Sophia University, Tokyo 102-8554, Japan}

\ead{m-akaki@sophia.ac.jp}

\begin{abstract}
We have investigated the magnetoelectric properties of (Ca$_{1-x}$Sr$_{x}$)$_2$CoSi$_2$O$_7$ ($0\leq~x\leq~1$) crystals with a quasi-two-dimensional structure. In Ca$_2$CoSi$_2$O$_7$ ($x=0$), a canted antiferromagnetic transition occurs at 5.6 K. The transition temperature  $T_{\rm N}$ is increasing with increasing Sr concentration, and the rises of the magnetization and dielectric constant become larger. Since the dielectric constant shows large change at  $T_{\rm N}$ and the magnetocapacitance effect is observed below  $T_{\rm N}$, a coupling between the magnetism and dielectricity is strong in (Ca$_{1-x}$Sr$_{x}$)$_2$CoSi$_2$O$_7$. The positive magnetocapacitance is reduced by Sr substitution, and is not observed in $x\geq 0.5$. Namely, the compound of $x\geq 0.5$ does not show the magnetic-field-induced electric polarization. On the other hand, the negative magnetocapacitance is enhanced by Sr substitution.
\end{abstract}

\section{Introduction}
Since the discovery of novel ferroelectricity due to spiral spin structures in TbMnO$_3$\cite{TbMn}, materials with spin frustration or nontrivial spin structures have attracted renewed interest as a promising candidate for new magnetoelectrics. 
The mechanism of the magnetic ferroelectricity is well explained by spin-current model\cite{Katsura} or exchange striction one\cite{Etype}.
According to spin-current model, the ferroelectricity originates in a spiral spin structure.
On the other hand, up-up-down-down spin order induces electric polarization through exchange striction.
However, the multiferroic materials that are not able to be explained by these models are recently discovered. 

Yi {\it et al}.\ have reported that Ba$_2$CoGe$_2$O$_7$ is such a new class of multiferroic materials\cite{BaCoGe}. In this context, we have focused on Ca$_2$CoSi$_2$O$_7$ and Sr$_2$CoSi$_2$O$_7$, which have the same crystal structure as Ba$_2$CoGe$_2$O$_7$: CoO$_4$ and SiO$_4$ tetrahedra are connected with each other through their corners to form two-dimensional layers, and the two-dimensional layers are stacking along the $c$ axis with intervening Ca or Sr layers (Fig.\ 1 (a), space group $P\overline{4}2_1 m$\cite{CaCo}).
Both compounds show the large magnetocapacitance, but their magnetoelectric properties are quite different from each other. 
In Ca$_2$CoSi$_2$O$_7$, electric polarization is induced perpendicular to the $c$ axis by applying magnetic fields parallel to the $c$ axis below canted antiferromagnetic transition temperature ($T_{\rm N}=5.6$ K). The large positive magnetocapacitance in Ca$_2$CoSi$_2$O$_7$ is caused by a magnetic-field-induced canted antiferromagnetic-paramagnetic transition\cite{AkakiAPL}.
On the other hand,  Sr$_2$CoSi$_2$O$_7$ does not show electric polarization even in magnetic fields. However, Sr$_2$CoSi$_2$O$_7$ show the large negative magnetocapacitance along the $c$ axis by applying magnetic fields perpendicular to the $c$ axis, and canted antiferromagnetic transition temperature ($T_{\rm N}=6.7$ K) is higher than Ca$_2$CoSi$_2$O$_7$\cite{AkakiCS}.
These results suggest that the mechanism of the large magnetocapacitance is different from each other.
For further understanding of the magnetoelectric properties of this system, we have systematically investigated the magnetoelectric properties of the series of (Ca$_{1-x}$Sr$_{x}$)$_2$CoSi$_2$O$_7$ ($0\leq x\leq 1$) single crystals.

\section{Experiment}
The single crystalline samples were grown by the floating zone method. We performed X-ray-diffraction and rocking curve measurements on the obtained crystals at room temperature, and confirmed that all samples have the tetragonal $P\overline{4}2_1 m$ structure without any impurity phases or any phase segregation.
The structural symmetry is unchanged over the whole Sr substitution range.
All specimens used in this study were cut along the crystallographic principal axes into a rectangular shape by means of X-ray back-reflection Laue technique. The magnetization and specific heat were measured using a commercial apparatus (Quantum Design, Physical Property Measurement System (PPMS)). The dielectric constant was measured at 100 kHz using an {\it LCR} meter (Agilent, 4284A).

\section{Results and Discussion}

\begin{figure}[tb]
\begin{center}
\includegraphics[width=0.98 \textwidth,clip]{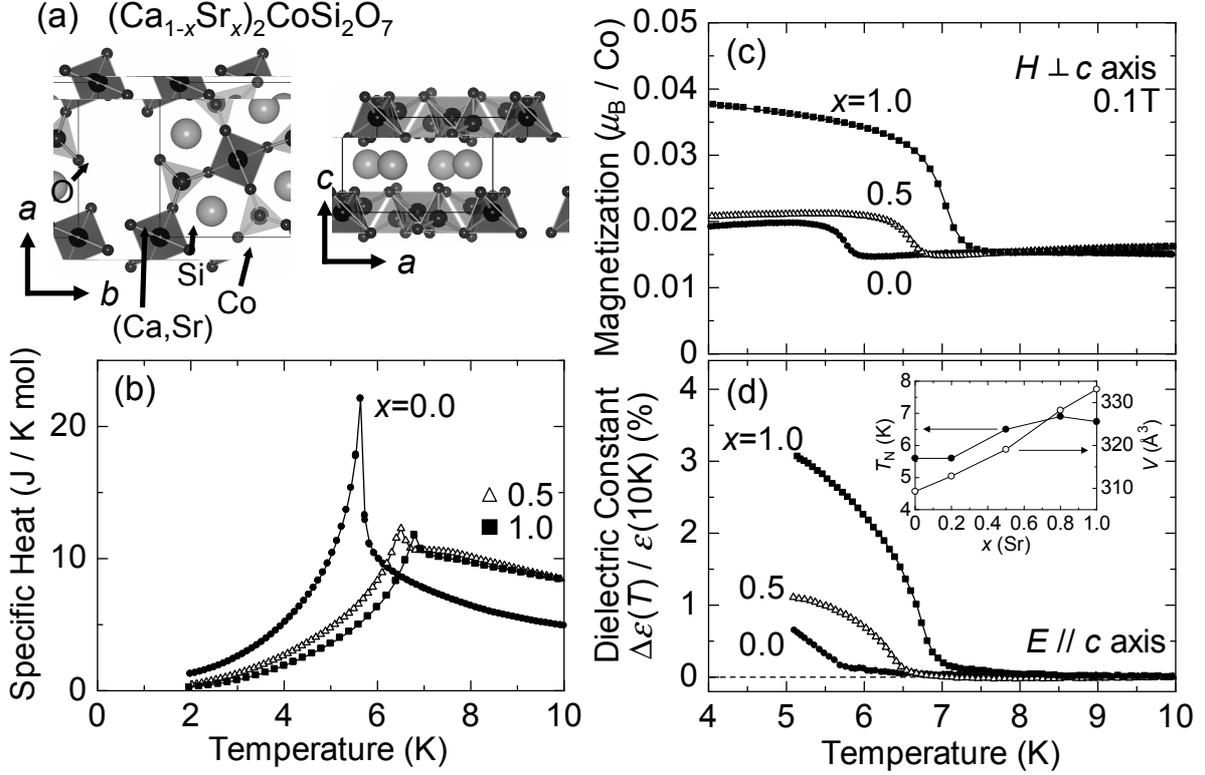}
\end{center}
\caption{(a) Schematic crystal structure of (Ca$_{1-x}$Sr$_{x}$)$_2$CoSi$_2$O$_7$ projected onto the $ab$ plane (left) and $ac$ plane (right). Temperature profiles of (b) specific heat, (c) magnetization, and (d) dielectric constant normalized by $\varepsilon ({\rm 10K})$ in (Ca$_{1-x}$Sr$_{x}$)$_2$CoSi$_2$O$_7$. The inset of panel (d) shows Sr concentration $x$ dependence of canted antiferromagnetic transition temperature $T_{\rm N}$ and lattice volume $V$\@.}
\label{f1}
\end{figure}


\begin{figure}[tb]
\begin{minipage}{0.50 \textwidth}
\begin{center}
\includegraphics[width=0.98 \textwidth,clip]{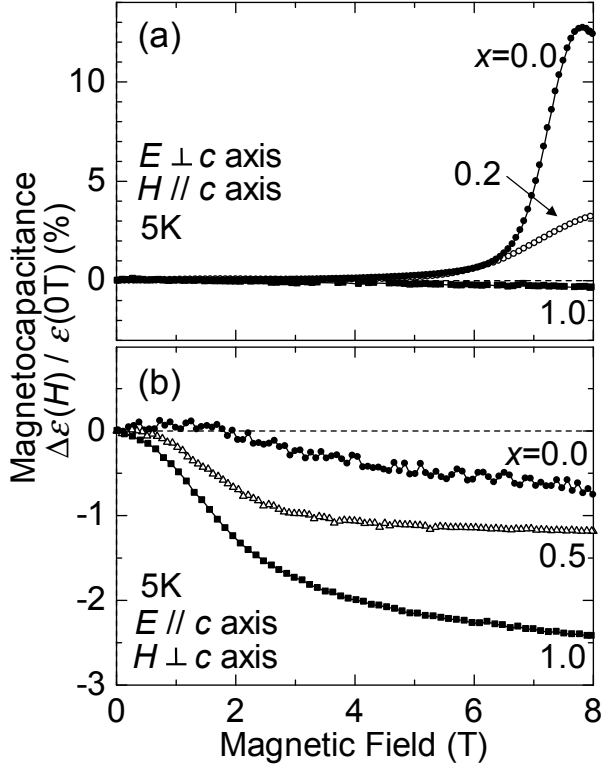}
\end{center}
\end{minipage}
\begin{minipage}{0.44 \textwidth}
\begin{center}
\caption{Normalized magnetocapacitance of (Ca$_{1-x}$Sr$_{x}$)$_2$CoSi$_2$O$_7$ crystals at 5 K as a function of an external magnetic field. (a) The dielectric constant perpendicular to the $c$ axis measured in magnetic fields along the $c$ axis. (b) The dielectric constant parallel to the $c$ axis measured in applied magnetic fields perpendicular to the $c$ axis.}
\label{f2}
\end{center}
\end{minipage}
\end{figure}

We present in Fig.\ 1 the temperature dependence of (b) specific heat, (c) magnetization, and (d) dielectric constant normalized by $\varepsilon ({\rm 10K})$ in (Ca$_{1-x}$Sr$_{x}$)$_2$CoSi$_2$O$_7$.
The magnetization shows a jump when applying magnetic fields perpendicular to the $c$ axis, indicating that a canted antiferromagnetic transition occurs at the temperature.
This magnetic transition does not accompany any thermal hysteresis, meaning that it is of second-order.
The dielectric constant parallel to the $c$ axis shows an abrupt increase below $T_{\rm N}$.
With increasing Sr concentration~$x$, $T_{\rm N}$ is increasing, and the rises of the magnetization and dielectric constant become larger.
Since the dielectric constant shows large change at $T_{\rm N}$, a coupling between the magnetism and dielectricity is strong in (Ca$_{1-x}$Sr$_{x}$)$_2$CoSi$_2$O$_7$.
As shown in the inset of Fig.\ 1 (d), the lattice volume is growing by Sr substitution. It is clear that the growth of the lattice volume is due to the larger ionic radius of Sr than Ca. 
The crystal structural changes as detected by the growth of the lattice volume enhance the antiferromagnetic superexchange. As a result, $T_{\rm N}$ increases.

Figures 2 shows the magnetic field dependence of the magnetocapacitance at 5 K\@. In (Ca$_{1-x}$Sr$_{x}$)$_2$CoSi$_2$O$_7$, the positive and/or negative magnetocapacitances are observed below $T_{\rm N}$, depending on the Sr concentration $x$ and the measurement configurations.
The positive magnetocapacitance is observed perpendicular to the $c$ axis for $x\leq 0.2$ when magnetic fields are applied parallel to the $c$ axis (Fig.\ 2 (a)), and electric polarization is induced.
This magnetocapacitance effect is caused by a magnetic-field-induced canted antiferromagnetic-paramagnetic transition\cite{AkakiAPL}.
The positive magnetocapacitance is reduced by Sr substitution, and not observed in $x\geq 0.5$. The compounds of $x\geq 0.5$ does not show the magnetic-field-induced electric polarization.
In Ca$_2$CoSi$_2$O$_7$ ($x=0$), structural phase transition between the commensurate and incommensurate phases is observed at 234 K much higher than $T_{\rm N}$\cite{CaCo2}.
It is probable that the anomalous crystal structure of Ca$_2$CoSi$_2$O$_7$ is closely linked to the appearance of the magnetic-field-induced electric polarization.
However, it is not observed in the compounds of $x\geq 0.5$.
These results suggest that the structural phase transition is significant for the magnetic-field-induced electric polarization of Ca$_2$CoSi$_2$O$_7$.
In contrast, with applying magnetic fields perpendicular to the $c$ axis, (Ca$_{1-x}$Sr$_{x}$)$_2$CoSi$_2$O$_7$ over the whole $x$ range shows the negative magnetocapacitance effect along the $c$ axis (Fig.\ 2 (b)). However, electric polarization is not observed for $x\geq 0.5$ even in this measurement configuration. The negative magnetocapacitance is enhanced by Sr substitution. 
The negative magnetocapacitance is due to the suppression of dielectric constant below $T_{\rm N}$ against magnetic fields. The scale of negative magnetocapacitance well corresponds with the variation of temperature dependence of dielectric constant in a zero magnetic field (Fig.\ 1 (d)).

\section{Conclusion}
In summary, we have investigated the magnetoelectric properties of (Ca$_{1-x}$Sr$_{x}$)$_2$CoSi$_2$O$_7$ ($0\leq~x\leq~1$) crystals and have observed the positive and/or negative magnetocapacitance effects depending on the Sr concentration $x$ and the measurement configurations. By the substitution of Sr for Ca, a canted antiferromagnetic transition temperature $T_{\rm N}$ is increasing, and the rises of the magnetization and dielectric constant become larger. The positive magnetocapacitance and magnetic-field-induced electric polarization are suppressed by Sr substitution, and is not observed in $x\geq 0.5$. The structural commensurate-incommensurate phase transition at much higher temperatures than $T_{\rm N}$ seems to be significant for the magnetic-field-induced electric polarization of (Ca$_{1-x}$Sr$_{x}$)$_2$CoSi$_2$O$_7$.
In contrast, Sr substitution enhances the negative magnetocapacitance arising from the different origin.

\section*{Acknowledgment}
This work was supported by Grant-in-Aid for JSPS Fellows and Scientific Research (C) from Japan Society for Promotion of Science.
\\

\end{document}